\documentclass{amsart}
\usepackage[utf8]{inputenc}
\usepackage{graphicx}
\usepackage{subcaption}

\usepackage{caption}
\usepackage{subcaption}
\usepackage{pgfplots}
\usepgfplotslibrary{groupplots}
\usepackage{amsmath,amssymb,amsfonts, amsthm}
\usepackage[foot]{amsaddr}
\usepackage{nicefrac}
\usepackage{bbm}
\usepackage{bm}

\usepackage{enumitem}
\usepackage[]{algorithm2e}
\usepackage{euscript}
\usepackage[bookmarks=true,hidelinks]{hyperref}
\numberwithin{equation}{section}		

\newcommand{\Dx}{{\ensuremath{\Delta}}}

\renewcommand{\phi}{\varphi}

\newcommand{\divx}[1]{\ensuremath{\nabla_x \cdot #1}}

\renewcommand{\vec}[1]{\ensuremath{\mathbf{#1}}}

\newcommand{\ddt}[1]{\ensuremath{\frac{\text{d}}{\text{d}t}#1}}
\newcommand{\hf}{{\ensuremath{\nicefrac{1}{2}}}}

\usepackage{subcaption}

\renewcommand{\geq}{\geqslant}

\def\XXint#1#2#3{{\setbox0=\hbox{$#1{#2#3}{\int}$ }
\vcenter{\hbox{$#2#3$ }}\kern-.56\wd0}}

\usepackage{todonotes}

\numberwithin{theorem}{section}

\theoremstyle{definition}

\usepackage{cleveref}

\newlength\figureheight
\newlength\figurewidth

\usepackage{booktabs}
\usepackage{algorithmicx}
\usepackage{algpseudocode}
\usepackage{graphicx}

\usepackage{etoolbox}

\newtoggle{usetikz}
\togglefalse{usetikz}
\newcommand{\InputImage}[3]{}
\iftoggle{usetikz}{%
	\renewcommand{\InputImage}[3]{%
		\setlength\figureheight{#2}%
		\setlength\figurewidth{#1}%
	}%
} { %
	\renewcommand{\InputImage}[3]{%
	}%
}
\pgfplotsset{every tick label/.append style={font=\tiny}}
\pgfplotsset{
	tick label style = {font = {\fontsize{6 pt}{12 pt}\selectfont}},
	label style = {font = {\fontsize{6 pt}{12 pt}\selectfont}},
	legend style = {font = {\fontsize{6 pt}{12 pt}\selectfont}},  
}

\usepackage{cleveref}
\usepackage[margin=2.5cm]{geometry}
\title{Alsvinn: A Fast multi-GPGPU finite volume solver with a strong emphasis on reproducibility}
\author{Kjetil Olsen Lye}
\address{Seminar for Applied Mathematics, ETH Z\"urich, R\"amistrasse 101, 8092 Z\"urich, Switzerland}
\email[Kjetil Olsen Lye]{kjetil.lye@sam.math.ethz.ch}
\date{November 2018}

\begin{document}
\begin{abstract}
	We present the Alsvinn simulator, a fast multi general purpose graphical processing unit (GPGPU) finite volume solver for hyperbolic conservation laws in multiple space dimensions. Alsvinn has native support for uncertainty quantifications, and exhibits excellent scaling on top tier compute clusters. 
\end{abstract}

\maketitle


\section{Introduction}
We are interested in approximating solutions to systems of non-linear hyperbolic conservation laws of the form
\begin{equation}
\left\{
\label{eq:conservation}
\begin{aligned}
    \vec{u}_t+\divx {\vec{F}(\vec{u})}&= 0\qquad \text{on } D\times [0,T]\\
    \vec{u}(x,0)&=\vec{u}_0(x)\qquad \text{on } D.
\end{aligned}\right.
\end{equation}
Examples include the compressible Euler equations of gas dynamics, the shallow water equations of oceanography, and the Magnetohydrodynamics equations of plasma physics. For a complete review, consult \cite{Dafermos}.

It is well-known that solutions of \eqref{eq:conservation} develop discontinuities in finite time, making the design of efficient numerical schemes challenging. The finite volume method has shown great success in dealing with these discontinuities, and will the method of choice for this paper. 

Additionally, instabilities, turbulence and multi-scale phenomena develop, and recent numerical and theoretical evidence~\cite{fkmt} show that deterministic solutions to \eqref{eq:conservation} are ill-posed, and one is forced to consider a probabilistic formulation. There are several popular frameworks available for uncertainty quantifications (UQ) for hyperbolic conservation laws, but due to the ill-posed nature of the equations, the  measure valued solutions~\cite{fkmt} and statistical solutions~\cite{FLM17,FLM18,FLMW19} are well suited for hyperbolic equations. 

The probabilistic formulations typically involve a sampling method, requiring multiple evaluations of the (very expensive) numerical scheme, which in turn creates the need for fast numerical solvers of hyperbolic conservation laws. While there are several multi-node CPU-based numerical finite volume codes, the largest computing clusters in the world often use general-purpose graphical processing units (GPGPUs) as their main accelerator, and thus one has a need for fast multi-GPGPU numerical codes for hyperbolic equations. Indeed, even highly optimized CPU codes are often limited by the lower memory bandwidth the CPU offers. Furthermore, since the intrinsic nature of the equations is probabilistic, the numerical solvers should have built-in support for uncertainty quantifications. On Switzerland's, and Europe's, largest supercomputer, the CSCS Piz Daint, the typical runtime difference of a highly optimized CPU finite volume code and an optimized GPGPU finite volume code can be an order of magnitude in favour of the GPGPU version. For three dimensional datasets, a single GPGPU does not have enough on-board memory to hold the whole dataset, and one is forced to develop multi-GPGPU finite volume solvers. While there are GPGPU finite volume solvers available, they are either often only available on single GPGPUs, or they do not have built-in support for UQ. Indeed, for high performance UQ on GPGPUs, one needs to take care that the statistics evaluation is done on GPGPU, minimzing overhead and storage requirements. 

Alsvinn, presented in this paper, is a fast, multi-GPGPU-based finite volume solver with UQ support built-in, written in C++ with CUDA and MPI, made to address the concerns in the above paragraph. Alsvinn supports a wide range of numerical finite volume solvers, and various sampling methods including Monte Carlo (MC), Quasi Monte Carlo (QMC) and Multilevel Monte Carlo (MLMC). Alsvinn is open source software freely available from \url{https://alsvinn.github.io}.

\section{Implementation of Alsvinn}

In this section we briefly describe the implementation and numerical methods used in Alsvinn. Alsvinn contains two core components: The finite volume solver, and the uncertainty quantification module. 


\subsection{Finite Volume Methods}

\label{subsec:fvm}
The finite volume method (FVM) is by far the most widely used method for approximating solutions of hyperbolic conservation laws. This section briefly reviews the finite volume method, for a full review, consult~\cite{leveque_green}.

We discretize the computational spatial domain as a collection of cells marked as $\vec{i}$, and we let $u^{\Dx}_{\vec{i}}(t)$ denote the averaged value in the cell at time $t\geq 0$. The semi-discrete case satisfies the following equation
\begin{equation}
\label{eq:semi_d}
\begin{aligned}
\ddt{}u^{\Dx}_{\vec{i}}(t)&+\sum_{k=1}^3\frac{1}{\Dx}\left(F^{k, \Dx}_{\vec{i}+\hf 
\vec{e}_k}(t)-F^{k, \Dx}_{\vec{i}-\hf 
\vec{e}_k}(t)\right)=0.
\end{aligned}
\end{equation}
Here $F^{k, \Dx}$ is a \emph{numerical flux function} in the direction $k$ for $k=1,2,3$. Alsvinn supports a large array of numerical fluxes, including the well-known HLLC flux\cite{hllc}, the Godunov flux, the Rusanov flux, the Roe flux, and high order entropy conservative fluxes described in~\cite{LeFloch2002}. We then use a strong stability preserving Runge-Kutta time stepping scheme~\cite{GST} to discretize~\eqref{eq:semi_d}, moreover, we also add a high order reconstruction, such as a the ENO or WENO reconstructions~\cite{Shu1999}, to estimate the cell interface values.

In Alsvinn, we use a smart mixture of compile-time and run-time polymorphism to support a multitude of equations, numerical fluxes, time steppers and reconstructions on both the CPU and the GPGPU. Moreover, the compile time polymorphism works akin to a compile time \emph{domain specific language}, allowing domain specialists to write their equations without interacting with any GPGPU specific code. Indeed, any of the mentioned components can be written once, but will automatically be translated into a version on the CPU and on the GPGPU. 

To minimize the overhead of internode communication when running in a multi-GPGPU setup, Alsvinn overlaps communication with computation, and both the communication and compactification of the halo domain is done in parallel with the computation of the inner domain.


\subsection{Uncertainty quantifications in Alsvinn}

Uncertainty quantifications are built into the Alsvinn simulator. We support a wide variety of sampling methods, both built in, but also user specified through external libraries. At the core, the UQ module of Alsvinn can approximate stochastic integrals through either single-level or multilevel sampling of the form
\[\int_{\Omega} G(u(\omega, \cdot))\; d\mathbb{P}(\Omega)\approx \frac{1}{M}\sum_{k=1}^M G(u_k) \qquad\text{and}\qquad \int_{\Omega} G(u(\omega, \cdot))\; d\mathbb{P}(\Omega)\approx \sum_{l=1}^L\frac{1}{M_l}\sum_{k=1}^{M_l}(G(u^l_k)-G(u^{l-1}_k)),\]
respectively. The random samples $1,\ldots, M$ are distributed over MPI by a load balancing algorithm.

We also support a wide range of statistical functionals, such as mean, variance, point PDFs, two-point structure functions, joint PDFs and multi-point correlators~\cite{FLM18}. And finally, a large variety of sampling methods are available, including Monte Carlo and Quasi Monte Carlo sampling. 

Alsvinn furthermore computes all statistics on the GPGPU when available, minimizing overhead, computational time and disk storage. Since each stochastic sample is independent, Alsvinn is able to perfectly parallelize over the stochastic dimensions for Monte Carlo and Quasi Monte Carlo,  while for MLMC, it can utilize effective load balancing algorithms. 

\subsection{Reproducibility}
Alsvinn is written with reproducibility in mind. It uses a range of modern software techniques to be as robust as possible. There are to date over 200 unit tests covering most of Alsvinn's functionality. To make the experiments easy to reproduce in other simulators, the input data is specified in an XML file specifying simulator parameters ( for example numerical flux and grid size), while the initial data is specified in a Python, both of which could potentially be read from any other finite volume simulator. 

Furthermore, every output file of Alsvinn is marked with the revision number of Alsvinn, the library versions used and the input parameters and initial data.


\section{Examples and performance analysis}
\label{sec:kh}
\begin{figure}
    \centering
    \begin{subfigure}{\textwidth}
    	    \centering
    \includegraphics[width=0.8\textwidth]{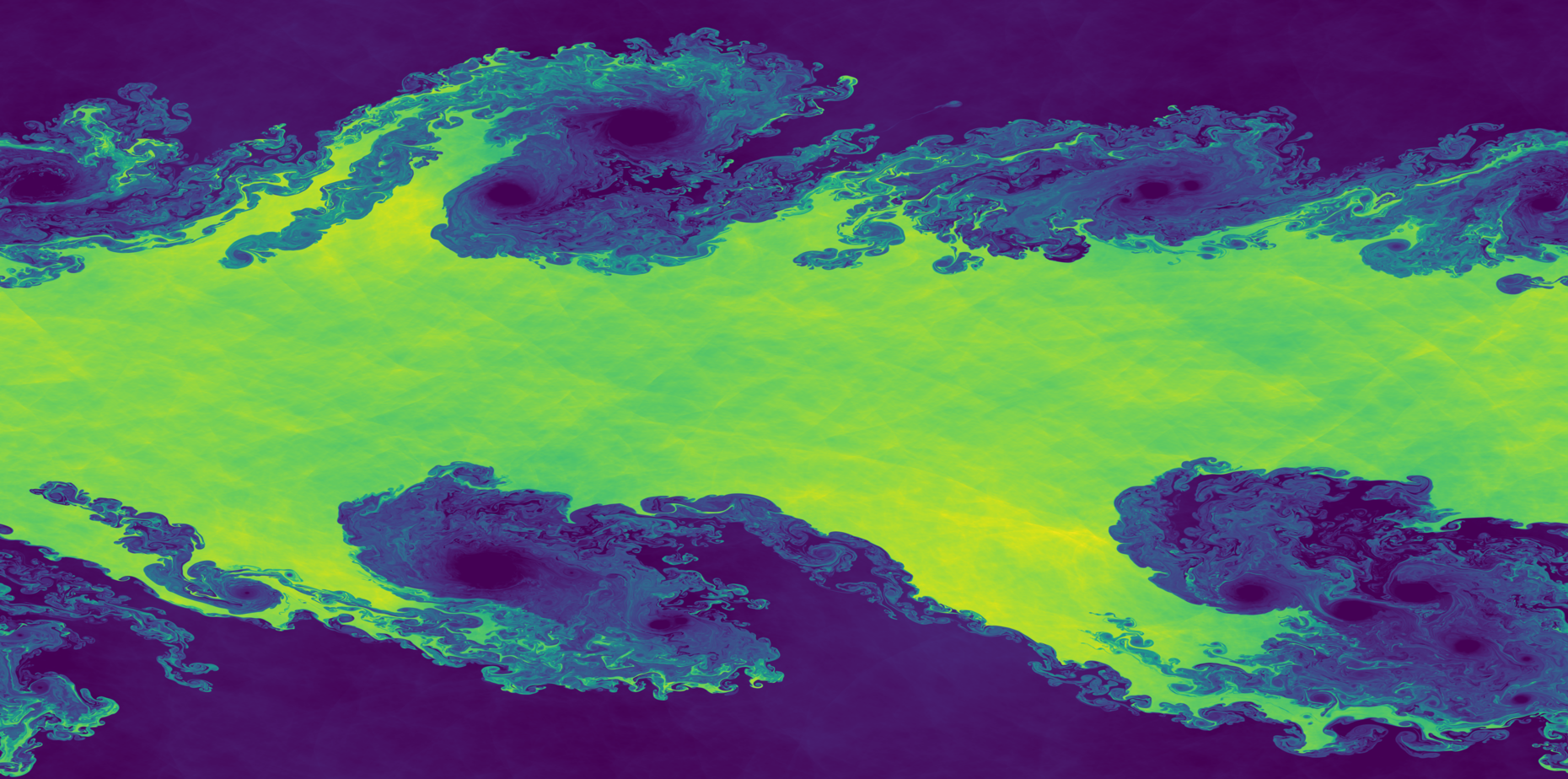}
    \subcaption{Single sample 2D}
    \end{subfigure}

    \begin{subfigure}{0.49\textwidth}
        	    \centering
    \includegraphics[width=0.8\textwidth]{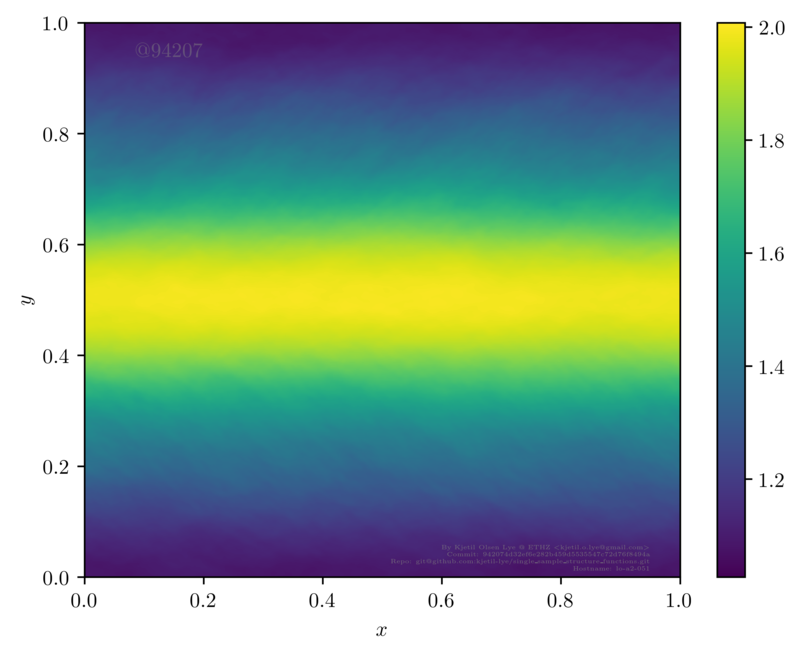}
    \subcaption{Mean 2D}
    \end{subfigure}
    \begin{subfigure}{0.49\textwidth}
    	    \centering
    \includegraphics[width=0.8\textwidth]{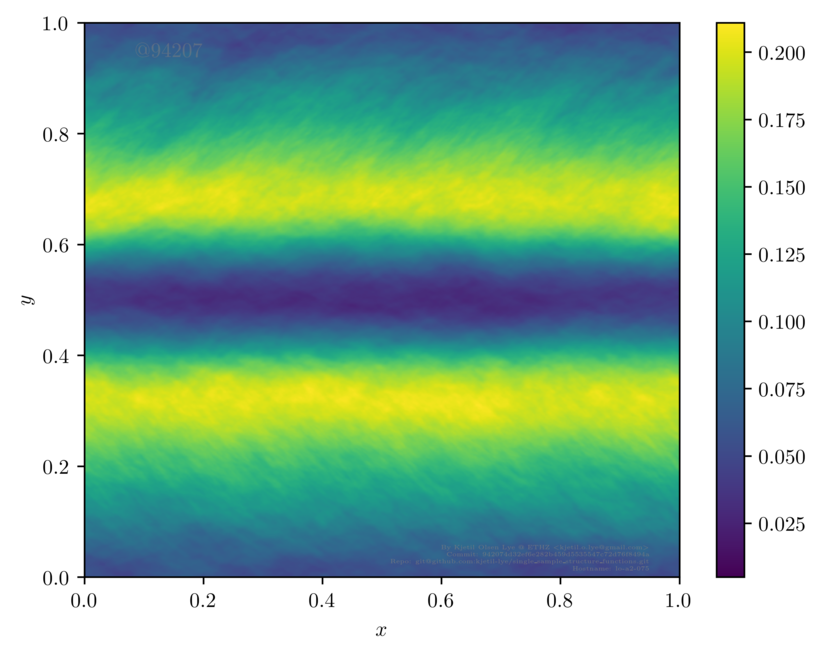}
    \subcaption{Variance 2D}
    \end{subfigure}

    \centering
    \begin{subfigure}{0.32\textwidth}
    	    \centering
    \includegraphics[width=0.8\textwidth]{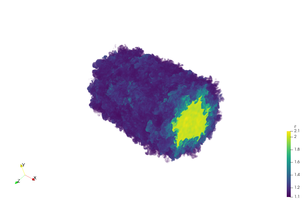}
    \subcaption{Single sample 3D.}
    \end{subfigure}
    \begin{subfigure}{0.32\textwidth}
    	    \centering
    \includegraphics[width=0.8\textwidth]{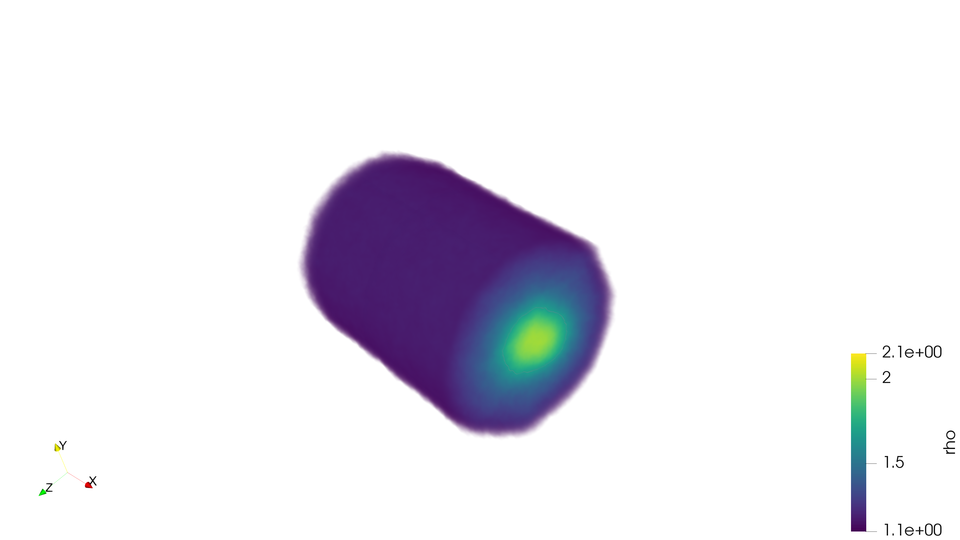}
    \subcaption{Mean 3D}
    \end{subfigure}
    \begin{subfigure}{0.32\textwidth}
    	    \centering
    \includegraphics[width=0.8\textwidth]{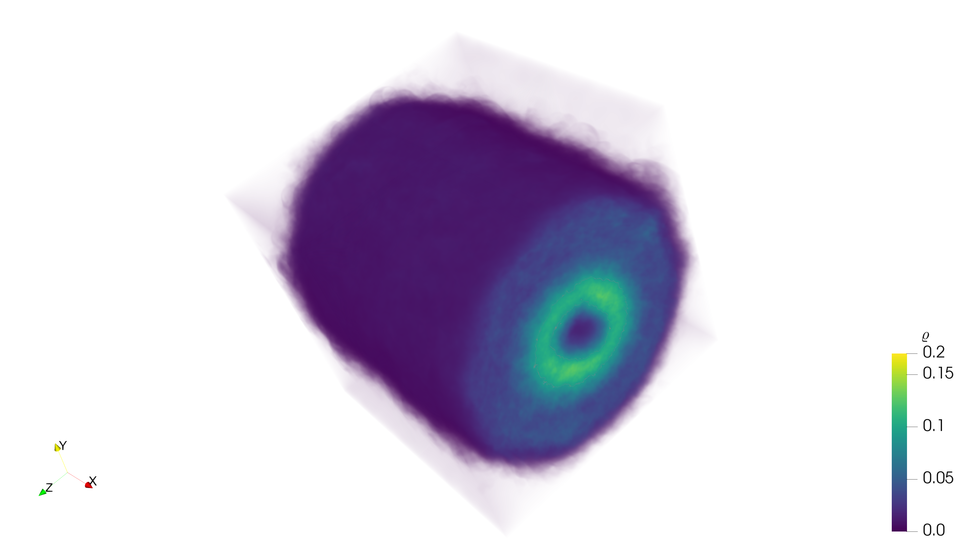}
    \subcaption{Variance 3D}
    \end{subfigure}
    
    \caption{Numerical results for the Kelvin--Helmholtz instability in 2D and 3D. We plot the approximate density ($\rho$) (A and E) in 2D and 3D, with a mesh resolution of $32768^2$ and $1024^3$ respectively, together with the mean and variance of the density (B, C, F and G). We also plot hte  The numerical flux used is the HLLC flux, a third order WENO reconstruction is used, along with a third order SSP Runge-Kutta time stepper \label{fig:kh_2d_3d}}
\end{figure}

The Kelvin--Helmholtz initial data is given as a shear flow with two states, see \cite{fkmt} for details. Here we use Alsvinn to simulate the Kelvin--Helmholtz initial data in two and three spatial dimensions, see \Cref{fig:kh_2d_3d} for an illustration.

\subsection{Spatial scalability}
We measure the performance of the communication implementation for 2 and 3 spatial dimensions for the Kelvin--Helmholtz experiment described in the above section. All runs are performed on the super computer CSCS Piz Daint.

\begin{figure}
    \centering
    \begin{subfigure}{0.45\textwidth}
    \includegraphics[height=0.2\textheight]{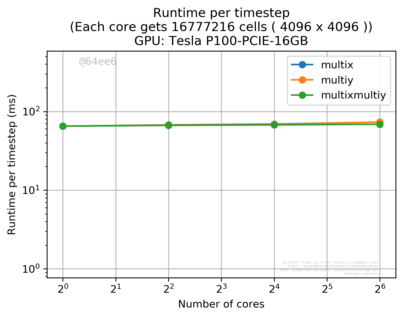}
    \subcaption{2D}
    \end{subfigure}
    \begin{subfigure}{0.45\textwidth}
    \centering
    \includegraphics[height=0.2\textheight]{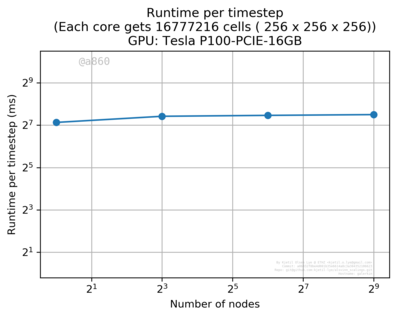}
    \subcaption{3D}
    \end{subfigure}
    
    \caption{\label{fig:kh_scaling}Weaking scaling of the Kelvin-Helmholtz experiment in 2D and 3D. In 2D, we can decompose either in the $x$-direction (multix), $y$-direction (multiy) or both (multixmultiy). In all experiments, an HLLC flux with a third order WENO reconstruction was used. Performed on CSCS Piz Daint. }
\end{figure}

 As parallel in time simulations of finite volume methods are still in its infancy, we focus our scalability study on the spatial decomposition. We perform two experiments to measure the weak scalability of Alsvinn. We run the Kelvin--Helmholtz experiment  2D and 3D with varying number of spatial cells, where we keep the number of cells per GPGPU constant (each GPGPU gets $2^24$ cells), and we measure the runtime per timestep in \Cref{fig:kh_scaling}. From both plots it is evident that Alsvinn obtains excellent scaling, and scales across 512 GPGPUs for a single sample. Here we note that even at a resolution of $2048^3$, we only use $0.105$ seconds per timestep.
 
\begin{figure}
    \centering
    \begin{subfigure}{0.45\textwidth}
    \includegraphics[height=0.2\textheight]{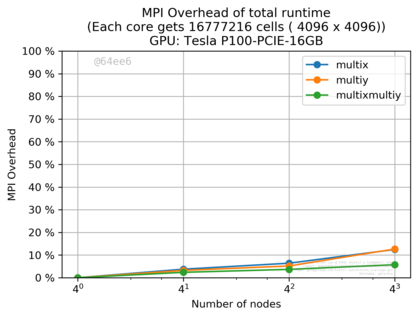}
    \subcaption{2D}
    \end{subfigure}
    \begin{subfigure}{0.45\textwidth}
    \centering
    \includegraphics[height=0.2\textheight]{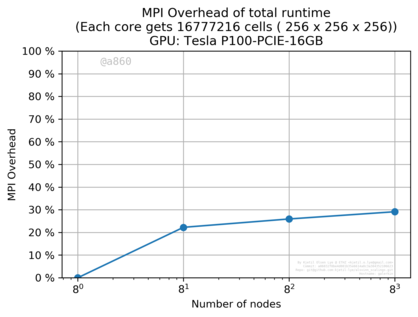}
    \subcaption{3D}
    \end{subfigure}
   
    \caption{ \label{fig:kh_overhead}MPI overhead for the Kelvin-Helmholtz experiment in 2D and 3D. In 2D, we can decompose either in the $x$-direction (multix), $y$-direction (multiy) or both (multixmultiy). In all experiments, an HLLC flux with a third order WENO reconstruction was used. Performed on CSCS Piz Daint. }
\end{figure}

 We furthermore measure the performance impact of the MPI communication compared to the overall runtime of the base configuration in \Cref{fig:kh_overhead}, that is, the MPI overhead of $K$ nodes is defined as
 \[\text{MPI Overhead}(K) = \frac{\text{Runtime per timestep}(K)-\text{Runtime per timestep }(1)}{\text{Runtime per timestep}(1)}.\] 
  As we can see, in 2D, we get less than 10 \% overhead, while in 3D, we get roughly 30\% overhead while utilizing 512 GPGPUs, which is to be expected for 3D GPGPU codes, where the dataset to communicate grows as $\mathcal{O}(N^2)$ where $N$ is the number of cells in each spatial direction.

\subsection{Utilization of hardware}
While the figures in the previous section showed that Alsvinn is able to scale over multiple compute nodes, it still remains to show that the code utilizes a single node effectively.  We benchmark our implementation on an NVIDIA Tesla P100. All experiments utilize double precision (64 bits) floating point numbers.

\begin{figure}
	\centering

	\begin{subfigure}{0.49\textwidth}
		\centering
		\includegraphics[width=\textwidth]{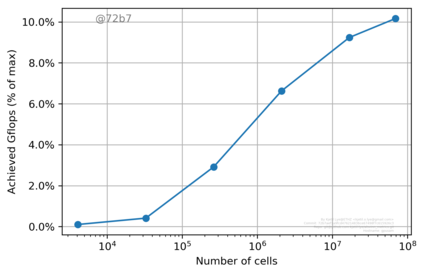}
		\caption{FLOPS, P100}
	\end{subfigure}
	\begin{subfigure}{0.49\textwidth}
		\centering
		\includegraphics[width=\textwidth]{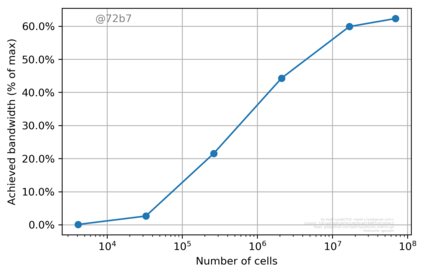}
		\caption{Bandwidth, P100}
	\end{subfigure}

	\caption{\label{fig:kh_p100_memory_bandwidth}Hardware utilization in Alsvinn. We measure the ratio between the achieved memory bandwidth and the maximum peak memory bandwidth (left), and the ratio between the achieved number of floating point operations per second (FLOPS) and the maximum number of floating point operations per second achievable by the GPGPU (right). In all experiments, an HLLC flux with a third order WENO reconstruction was used. Performed on CSCS Piz Daint with an NVIDIA Tesla P100.}
\end{figure}

We plot the a ratio between the achieved number of floating point operations per second (FLOPS) and the maximum FLOPS achievable on the GPGPU in~\Cref{fig:kh_p100_memory_bandwidth} (A). We see that we achieve 10 \% of peak floating point performance, which is to be expected from a finite volume algorithm, which usually is memory bandwidth bounded.

In \Cref{fig:kh_p100_memory_bandwidth} (B) we measure the ratio between the achieved memory bandwidth, and the theoretical maximum memory bandwidth of the GPGPU. From the figure, we see that we achieve above 60 \% memory bandwidth utilization, which confirms that the code is memory bandwidth bounded, and that we achieve close to peak performance in terms of the memory bandwidth.

\section{Conclusion}
We have presented Alsvinn, a robust, fast, efficient and highly scalable multi-GPGPU code for simulating and performing UQ for systems of conservation laws, such as the Euler equations of gas dynamics, in two and three space dimensions. It has been well-established \cite{fkmt} that purely deterministic simulations of multi-dimensional conservation laws may not converge on mesh refinement. Hence, UQ simulations have to be performed even of deterministic initial data. Given the prohibitive cost of simulating complex three-dimensional problems with state of the art multi-CPU codes on leadership class hardware platfroms, it is imperative to design scale multi-GPGPU codes. Alsvinn satisfies these requirements as it is shown to scale very well on state of the art GPGPU clusters and incorporates various efficient numerical schemes and sampling based UQ techniques in one platform. In particular, we are able to perform UQ, within the framework of statistical solutions, for very complex three dimensional compressible flows at high spatio-temporal and statistical resolutions, with Alsvinn, providing the first such results globally. 

\bibliographystyle{plain}
\bibliography{biblo}{}


\end{document}